# A review of knowledge graph application scenarios in cyber security

Kai Liu[1], Fei Wang[1], Zhaoyun Ding[1], Sheng Liang[2], Zhengfei Yu[1], Yun Zhou[1]

[1] Science and Technology on Information Systems Engineering Laboratory, National University of Defense Technology, China
[2] Center for Information and Language Processing (CIS), University of Munich (LMU), Germany

Corresponding author(s). E-mail(s): liukai18@nudt.edu.cn, zyding@nudt.edu.cn

## Abstract

Facing the dynamic complex cyber environments, internal and external cyber threat intelligence, and the increasing risk of cyber-attack, knowledge graphs show great application potential in the cyber security area because of their capabilities in knowledge aggregation, representation, management, and reasoning. However, while most research has focused on how to develop a complete knowledge graph, it remains unclear how to apply the knowledge graph to solve industrial real challenges in cyber-attack and defense scenarios. In this review, we provide a brief overview of the basic concepts, schema, and construction approaches for the cyber security knowledge graph. To facilitate future research on cyber security knowledge graphs, we also present a curated collection of datasets and open-source libraries on the knowledge construction and information extraction task. In the major part of this article, we conduct a comparative review of the different works that elaborate on the recent progress in the application scenarios of cyber security knowledge graph. Furthermore, a novel comprehensive classification framework is created to describe the connected works from nine primary categories and eighteen subcategories. Finally, we have a thorough outlook on several promising research directions based on the discussion of existing research flaws.

**Keywords**： Cyber security, Knowledge graph, Construction technology, Application scenarios

## 1. Introduction

With the development of new information technologies and applications, the scale of the cyber space is gradually expanding from the traditional internet to a variety of areas such as manufacturing, healthcare, agriculture, aviation, business, etc. As a result, cyber space can comprise interactions between industrial physical systems, human social systems, and network information systems and has become an increasingly complex infrastructure for social development. The opportunities left for attackers are increasing. Due to their combination of cyber as well as many physical assets, the consequence of cyber-attacks become more and more serious. The cyberattack experienced by Colonial Pipeline is an example of how a cyberattack can impact the physical world. The cyberattack shut down a pipeline that supplies 45% of the East Coast's fuel, leading to a $5 million economic loss，fuel delivery disruption, and panic buying across the United States [1]. Given the increasing number and intensity of attacks and malware, the lack of qualified cyber security personnel is a cause for concern [2]. Since neither the number of available people nor the required skills can be increased overnight, companies must increase the development of technologies for modeling experts' knowledge and experience. The integration of automation, intelligent technology, and attack defense technology has become one of the inevitable trends in the development of cyber security technology.

Cyberattacks and defenses against them are conducted in dynamic complex environments, with numerous factors contributing to attack success and mission impacts. The network environments are continually changing, with the applications installed, machines added and removed, etc., which is one of the main obstacles [3]. On the other hand, the information asymmetry between the offensive and defensive sides in cyberspace is becoming more and more obvious [4]. For example, when confronted with a constantly updated new vulnerability or attack pattern, defenders often feel helpless to grasp the most up-to-date attacking techniques and vulnerability information, as well as the corresponding effective defense strategy, to maintain a balance with the attacker. The long persistence and highly concealed characteristics of modern attacks, such as advanced persistent threat (APT) attacks [5], make the limitations of traditional defense technologies based on expert rules, machine learning, and deep learning have become increasingly apparent. The relatively simple tasks, such as feature extraction [6], anomaly detection [7], and data classification [8], can no longer restore the full picture of attack behavior. Expert knowledge hidden in cybersecurity data is still a very important breakthrough to solve the above problems.

However, the cyber security-related data generated in cyberspace has experienced explosive growth. These data are diverse, heterogeneous, and fragmented, making it difficult for cyber security managers to quickly find the information they need [9]. Therefore, the current problem of cyber security analysis is not the lack of available information, but how to assemble heterogeneous information from multiple sources into one model, to further understand the cybersecurity situation and provide auxiliary decision support. The current focus of cybersecurity analysis research is to obtain correlations and potential attacks from threat intelligence data. Technologies such as correlation analysis [10], causal inference [11], and semantic reasoning [12] technologies based on knowledge modeling have become new solutions under big data conditions.

The cybersecurity knowledge graph (CSKG), as a specific knowledge graph (KG) in the security area, is made up of nodes and edges that constitute a large-scale security semantic network, providing an intuitive modeling method for various attacks and defense scenarios in the real security world. Nodes can be entities or abstract concepts (e.g., vulnerability name, attack pattern, product name, vendor), edges represent the attributes or relationships between entities, and nodes and edges together form a KG. The advantages of the KG can be discussed under three aspects: first, utilizing KG construction and refining techniques including ontology [13], information extraction (IE) [14,15], and entity disambiguation [16], KGs effectively extract and integrate existing knowledge from multi-source heterogeneous data. Second, it can express knowledge in the cyber security domain structurally and relationally, and visualize the knowledge in a graphical manner, which is very intuitive and efficient. Third, using semantic modeling, query, and reasoning technologies, a cyber security KG can imitate the thinking process of security specialists aims to derive new knowledge (as known as new relations) or check data consistency based on the existing facts (i.e., triples) and logic rules [17].

Although construction of CSKG recently caught the attention of both researchers and companies. Many kinds of CSKGs were constructed from different perspectives of cyber security. However, while most research has focused on how to develop a complete KG, it remains unclear how to apply the KG to solve industrial real challenges in cyber-attack and defense scenarios. Some research teams have proposed some schemas and made some attempts, but there is still a long way to go before they can be put into practice. How, for example, should KGs be used in a specific network environment with specific network assets? Security managers in many organizations wonder whether the existing CSKG can be reused in their work and whether it fits into their existing IT infrastructure. Furthermore, few earlier papers addressed what kind of new knowledge the CSKG may infer in addition to new relationships.

Previous systematic reviews or meta-analyses of the CSKG have been mainly undertaken in its data processing, construction, and visualization. Kai Zhang [18] etc. reviewed the literature about the application of KG only from the aspects of situation awareness, security assessment and analysis, and association analysis, which was limited to the security assessment region. Noel [19] summarized the graph-based methods for assessing and improving network security in two major aspects: the "when" aspect and the "where" aspect. The first-dimension cover three particular phases of the security process (prevention, detection, or reaction). In the second dimension, it gave an expectation that incorporating various operational components (i.e., network infrastructure, security posture, cyber-space threats, mission dependencies) into a unified knowledge base for many cyber security tasks. The other articles reviewed the research of CSKG mainly from the dimensions of data sources, ontology design [20], construction technologies [21], and reasoning methods [22]. As a part of the review, Ding et al. [23] briefly attempted to illustrate several application directions of CSKG based on the introduction of CSKG construction technologies. However, none of the survey papers mentioned above focused on the challenging problem that how to utilize the CSKGs to solve practical issues?

The goal of this paper is to motivate and give an introduction to the application scenario of KGs in cyber security. To have a comprehensive survey of current literature, this paper first gives a summary of the background and construction methods of the CSKG. The primary content of this paper focuses on providing an overview of existing application scenarios of CSKGs and related datasets found in practice. Finally, we discuss the future directions of this research field. Our main works are as follows:

- **A comprehensive review of existing application scenarios of CSKG**. We propose a novel classification framework for conducting a comprehensive review of the application scenarios of CSKG based on an investigation of the background and construction technology of CSKG.
- **We Summarize the relevant datasets**. To facilitate future research on CSKGs, we also provide and analysis a curated collection of datasets and open-source libraries for two tasks: the construction task of CSKG and the task of information extraction.
- **The future directions**. This survey provides a summary of each category and highlights promising future research directions.

The rest of this paper is organized as follows. **In Section 2**, an overview of construction methods for CSKG including definitions, the building flow, ontology, named entity recognition methods, and relation extraction approaches are given. The usual datasets, as well as their inadequacies, are presented in section 3 to aid in the application of CSKG and the extraction of information. In **Section 4**, an overview of the application progress of KGs in the cyber security domain is given. In s**ection 5**, we discuss the shortcomings of existing research before prospecting future research opportunities. Finally, we conclude the paper in **Section 6**.

## 2. A brief overview of construction methods for cyber security knowledge graph

Knowledge representation has experienced a long-period history of development in the fields of logic and AI. With resource description framework (RDF) [24] and Web Ontology Language (OWL) [25] were released in turn, the idea of modern KG gained great popularity since its first launch by Google's search engine [26] to enhance the search engines capabilities and the user's search quality. In this section, we will briefly introduce the construction process of CSKG from the following aspects: First, we present a framework for building a CSKG. Then, the security ontology design is described as representing knowledge in the security domain. Next, our review goes to tasks of named entity recognition. Finally, we talk about relation extraction and similar works in this domain.

## 2.1 Some definition

KGs are structured semantic knowledge bases used to describe concepts and their interrelationships in the physical world in the symbolic form [27]. Formally, a KG can be typically defined as $G = (E, R, T)$, where $G$ is a labeled and directed multi-graph, and $E = \{e_1, e_2, \cdots e_{|E|}\}$, $R = \{r_1, r_2, \cdots r_{|R|}\}$ are the set of entities and relationships respectively. $|E|$ and $|R|$ represent the number of elements in sets E and R, respectively. Each triple is formalized as $T = \{(e, r, e') \mid e, e' \in E, r \in R\}$ which represents a fact of the relation $r$ from head entity $e$ to tail entity $e'$. The triplet $< entity, relationship, entity >$ and $< concept, attribute, value >$ are the basic forms of $T$, in the knowledge base. The entities as the basic elements in the KG mainly include collections, categories, object types, categories of things (e.g., product, vendor, vulnerability, attacker). Relationships connect entities to form a graph structure, and the attribute contains characteristics and parameters, such as google.com, windows, etc.

## 2.2 The building flow of KG

Similar to the general KG construction process, the CSKG follows the process and framework of the general KG construction. Due to the relatively mature and complete knowledge data of this field, a top-down construction method [28] can be adopted for building CSKG. The fragmented domain data could be integrated under the guidance of a certain framework or a pre-designed cyber security ontology from domain experts. Then information extraction and entity alignment technologies can separate entities and relationships from the original cyber security data. Knowledge reasoning technology can generate new knowledge based on the existing KG to provide support for prediction and inference tasks. The construction framework of the CSKG is shown in **Figure 1**.

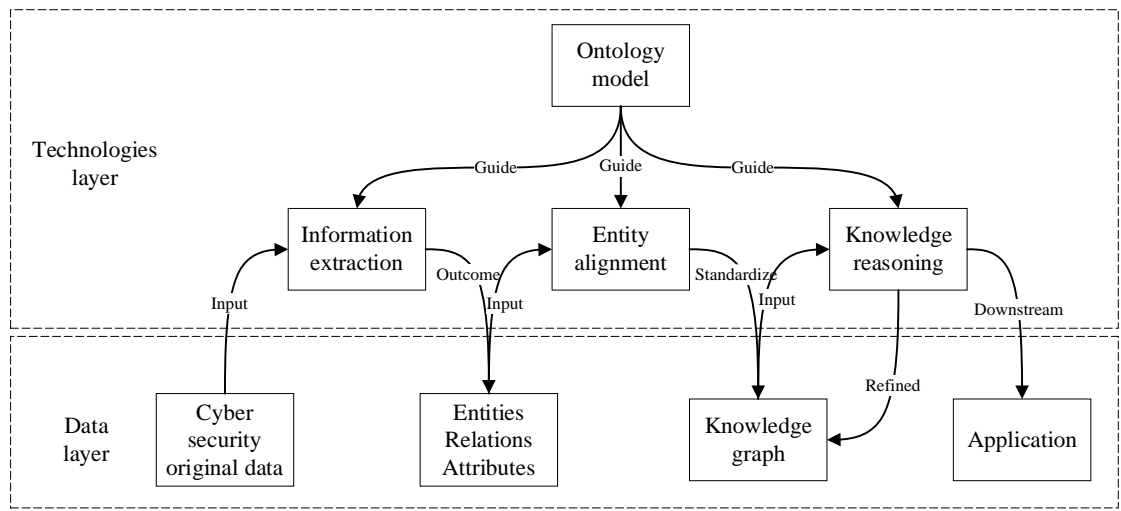

**Figure 1 The construction framework of the cyber security knowledge graph**

## 2.3 Cyber security ontology

Cyber security ontology is used to describe cyber security concepts and relationships between concepts in a cyber security field or even a wider range. These concepts and relationships have a common, unambiguous, and unique definition that everyone agrees on in the shared range, which makes humans and machines can communicate with each other [29]. Unified ontologies, such as STUCCO [30], Unified Cybersecurity Ontology (UCO) [31], were created in the field of cyber security to incorporate and integrate heterogeneous data and knowledge schemas from various cybersecurity systems, as well as the most commonly used cybersecurity standards for information sharing and exchange. For different specific application scenarios, researchers have developed different ontologies, such as intrusion detection [32], malware categorization [33]and behavior modeling [34], cyber threat intelligence (CTI) analysis [35], cyber-attack analysis [36, 37], threat and security evaluation [38, 39], vulnerability analysis [40], threat actor analysis [41], etc. as shown in **Figure 2**. Building a generic network security ontology in today's complex cyber environment is a difficult and time-consuming process that heavily relies on the domain knowledge and information technology knowledge of network security professionals. As a result, application scenarios should guide the design of appropriate security ontology. At the same time, dynamic and automatic enrichment of the information security ontology is required [42].

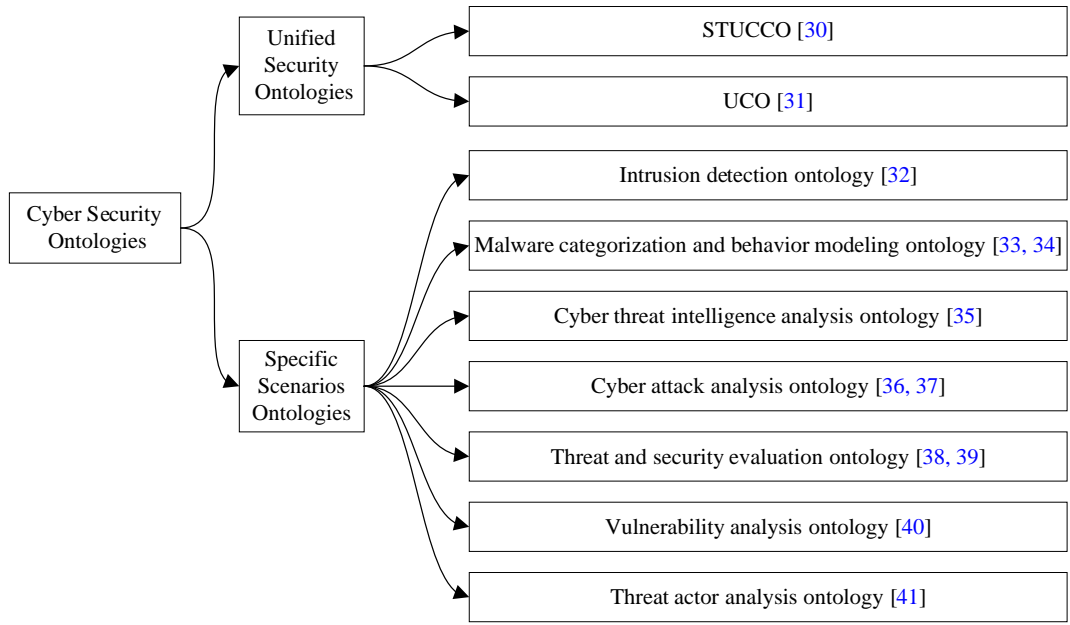

**Figure 2 The cyber security ontologies**

## 2.4 Cyber security entities extraction

The technology of information extraction (IE) has drawn incremental attention. At present, there are two main tasks of IE are Named Entity Recognition (NER) and Relation Extraction (RE). Traditional approaches to NER are broadly classified into three main streams: rule-based, unsupervised learning and feature-based supervised learning approaches [43]. The rule-based methods, such as regular expression [44], bootstrapping methods [45], etc. work very well when the lexicon is exhaustive, but cannot be transferred to other domains. Comparatively, traditional statistical-based extraction methods [46] including the Hidden Markov model (HMM), Decision Trees, Maximum Entropy Model (MEM), Support Vector Machines (SVM), and Conditional Random Fields (CRF), achieve good results. However, they rely heavily on feature engineering, which has some limitations [47]. Compared to traditional approaches, deep learning is beneficial in the capability of representation learning and the semantic composition empowered by both vector representation and neural processing. Illustrated in **Figure 3** are the three basic components (i.e., distributed representation, feature extractor, and decoder) and some corresponding instances of the deep learning NER approach.

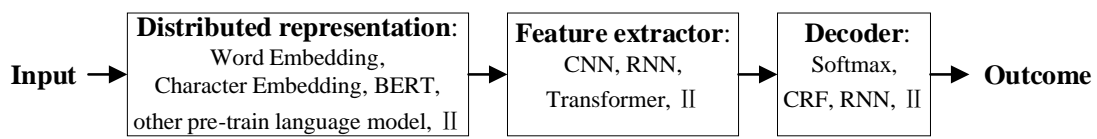

**Figure 3 The deep learning approach for the NER flowchart**

This allows a machine to be fed with raw data and to automatically discover latent representations and processing needed for classification or detection. At present, many methods [48-

52] have been tried, mainly including multi-task deep learning, deep transfer learning, deep active learning, deep reinforcement learning, deep adversarial learning, attention mechanism, etc.

## 2.5 Relations extraction of cyber security entities

The relationship between entities is an indispensable part of the KG. Abundant relationships weave independent entities together into a KG. Relationship extraction from unstructured text is one of the core tasks of the KG construction. To overcome the limitation of traditional methods that traditional ways strongly rely on the quality of hand-engineered features, Zeng et al. [53] proposed an end-to-end CNN-based method that could automatically capture relevant lexical and sentence-level features. The methods based on RNN or LSTM have been proposed one after another [54, 55]. However, most supervised relation extraction methods require large labeled training data which is expensive to construct. Distant Supervision (DS) [56] helps with the construction of this dataset automatically. Furthermore, multi-instance learning, sentence-level attention mechanism [57, 58] were leveraged in the task of relation extraction to reduce the noise introduced by DS. The above-mentioned work typically solves IE tasks in the extract-then-classify or unified labeling manner. However, these methods either suffer from redundant entity pairs or ignore the important inner structure in the process of extracting entities and relations. To address these limitations, Fu et al. [59] presented a method of joint extraction of entities and relations, which outperformed the previous pipelined approach. Some researchers attempt to extract cyber security entity-relation triples using a joint extraction model based on their unpublished datasets, which does not address the issue of labeled data scarcity [60, 61].

However, there are three main challenges during extracting information from unstructured cybersecurity text. Firstly, most previous IE research has focused on common events in a person's life, such as those defined by ACE [62] or the TAC Knowledge Base Population [63]. One core difference between extracting life knowledge and cybersecurity knowledge is the required domain-relevant expertise. As a result, the IE task suffers from a lack of large labeled training data. A second difference between extracting life knowledge and cybersecurity knowledge is the inherent complexity of cybersecurity knowledge. A cyberattack event can consist of an attack pattern with multiple actions, attempted or completed. Each mention of one of these actions can be considered as a separate cybersecurity event description, which multiplies the possible choices for a cybersecurity event reference. Thirdly, there is much implicit information in the unstructured data that cannot be expressed explicitly.

To assure the quality of the KG, it is required to analyze and verify the knowledge, remove redundant knowledge, and resolve conflicting information to prevent errors from propagating in the reasoning process. The technologies, such as named entity linking and entity disambiguation [64], were proposed to refine the KG. Additionally, the KG, which is generated based on IE technologies, primarily contains relationships that are represented in sentences explicitly. It is also required to mine possible implicit knowledge and enrich the CSKG through reasoning [65]. The knowledge reasoning may be combined with specific task requirements, and it can employ association query, rule-based reasoning, distributed representation learning-based reasoning methods, and neural network-based reasoning methods extensively [66].

# 3. The datasets

In their mission to secure systems, security analysts rely on a wealth of knowledge such as known and newly identified vulnerabilities, weaknesses, threats, and attack patterns. Such knowledge is collected, published, and structured by research institutions, government agencies, and industry experts, e.g., Computer Emergency Response Teams (CERTs) and MITRE [67]. The widely used standards include the vulnerabilities and associated data from the National Vulnerability Database (CVE, CWE, CPE, CVSS, etc.) [68], and potential attacker exploits expressed with Common Attack Pattern Enumeration and Classification (CAPEC) [69]. In this section, we review the significant datasets for building CSKG by the categories as follows: (1) the datasets of open-source CSKG; (2) the datasets for IE in the cyber security domain; (3) the other datasets, which may enlighten new researchers to find new solutions.

## 3.1 The datasets of open-source CSKG

Table 1 presents a comparative study of the datasets for CSKG presented. For each dataset, the purpose, data sources, and relevant papers (if available) are considered. These open-source CSKG are developed for different purposes, which means they need various data sources. For instance, the CWE-KG [72] was constructed to discover potential threats from Twitter data, so it needs the information from CWE, CAPEC, and Twitter data. There are, however, obvious drawbacks to these datasets. One of these is that four of them do not give the statement document, e.g., paper, report. More than this, the Vulnerability KG [75] even only displays the visualization results on a webpage, without the statement of its construction method and experiment performance. The Open-CyKG [76] presents a CTI-KG framework that is constructed using an attention-based neural Open IE model to extract valuable cyber threat information from unstructured APT reports. Moreover, the MalKG [79] aims to integrate information on malware threat intelligence. The SEPSES CKB [70] provides the details of the CSKG dataset and the corresponding published paper [71]. While the knowledge from MITRE provided strong support for building this KG, this study neglect to include the open-source cyber threat intelligence (OSCTI). Furthermore, these works are also limited by their only consideration of one language. Thereby, the researchers in the industry are beginning to look forward to the development of CSKG based on abundant threat intelligence and more languages.

**Table 1: Comparative table of the datasets of open-source CSKG**

| Name and Year | Purpose | Data sources | Ref. |
|---|---|---|---|
| SEPSES CKB [70], 2019 | SEPSES CSKG with detailed instance data | CVE, CWE, CAPEC, CPE, CVSS. | [71] |
| CWE-KG [72], 2019 | A CWE KG supporting Twitter data analysis | CWE, CAPEC, Twitter data | - |
| CSKG [73], 2020 | Cybersecurity KG | CVE, CWE, CAPEC | - |
| ICSKG [74], 2020 | KG for vulnerabilities of industrial control systems | CVE, CWE, CPE, CERT | - |
| Vulnerability KG [75], 2021 | Visualization web page of vulnerability KG | CVE, CWE | - |
| Open-CyKG [76], 2021 | An open cyber threat intelligence KG | APT reports, CTI reports | [77] |
| MalKG [78], 2021 | Open-source KG for malware threat intelligence | CVE, Malware reports | [79] |

## 3.2 The datasets for information extraction task

Information extraction technologies are essential for generating CSKG. For the purpose of training robust IE models, providing quality-assurance annotated datasets is a crucial task that cannot be bypassed. This chapter summarizes several datasets for enhancing the works for knowledge extraction. From the discussion above it is apparent that the datasets should be divided into two categories: the datasets for the NER task and the datasets for the RE task. Frankly speaking, the research around this topic is more than these, but the others did not publish their datasets. The datasets listed in Table 2 are only the part of new researchers could download and use. Despite the fact that most of the datasets are defined for the NER task, following the trend that Rastogi et al .'s new malware dataset [79] is the only one for both the NER and RE tasks. The entity and relation types are in general defined in a specific security ontology, which restricts the potential for use outside of the designated domain. Although they are collected and annotated based on various data sources, most of them are generated with English corpus. On the other hand, a lot of security knowledge is released in a variety of languages, often intermingled with English, necessitating the creation of multi-language datasets.

Table 2: The open datasets of information extraction task for cyber security

| Dataset and Year | Task | | Entity types | Data sources |
| --- | --- | --- | --- | --- |
| | NER | RE | | |
| Lal [80], 2013 | √ | - | Software, Network_terms, Attack, File_name, Hardware, Other_technical_terms, NER_modifier | Blogs, Official Security Bulletins, CVE |
| Bridges et al. [81], 2013 | √ | - | Vendor, Product, Version, Language, vulnerability, and vulnerability relevant term | NVD, OSVBD, Exploit DB |
| Lim et al. [82], 2017 | √ | - | Action, Subject, Object, Modifier | APT reports, |
| Kim et al. [83], 2020 | √ | - | Malware, IP, Domain/URL, Hash, their categories | CTI reports |
| Rastogi et al. [79], 2021 | √ | √ | Malware, MalwareFamily, Attacker, AttackerGroup, ExploitTarget, Indicator, etc. | CVE, Malware reports |

## 3.3 Other datasets for the application of CSKG

Transferring theoretical framework to practice by mastery of the environment, understanding of the actions of threat actors, integration of external intelligence, and stockpiling of basic knowledge. In 2021, Zhang et al. [84] categorized the data demand from four dimensions: environment data (e.g., assets and their weakness), behavior data (such as network alerts, terminal alerts, and logs), threat intelligence (e.g., internal and external CTI), and knowledge data (such as ATT&CK and CAPEC). However, the current studies highlight the threat intelligence and the knowledge data in building CSKG and ignore the environment and behavior data of the target IT system. Further, there is still no mature and unified specification to describe these data.

# 4. The application scenarios

After being introduced by Google, knowledge graph technology has attracted a lot of research interest in recent years. In the cyber security domain, research on KG can be classified into two categories, research on construction techniques and applications. Studies on construction techniques focus on the extraction, representation, fusion, and reasoning of the knowledge in the graphs [85], such as linking entities and relations to KG correctly after extracting them from unstructured text and reasoning new facts from such KG. While studies on application stress applying CSKG to solve practical problems in different network environments. This paper gives a systematic survey on the applications of CSKGs.

According to our current survey, the majority of papers devoting to applying KGs into specific areas have put their interests in different particular phrases of overall security process (assessing the overall situation of the network, discovering potential threats, and investigating the ongoing or ending attack), which will be introduced in from section 4.1 to section 4.3 in this paper. This article presents several specific applications from section 4.4 to section 4.8 that will help operators and managers with decision making, operation, vulnerability management, malware attribution, and analysis in conjunction with the physical environment. Section 4.9 introduces some other possibilities of application for CSKG like social engineering attack analysis, fake cyber threat intelligence identification. A taxonomy of the application of CSKG in this paper is given in **Figure 4**.

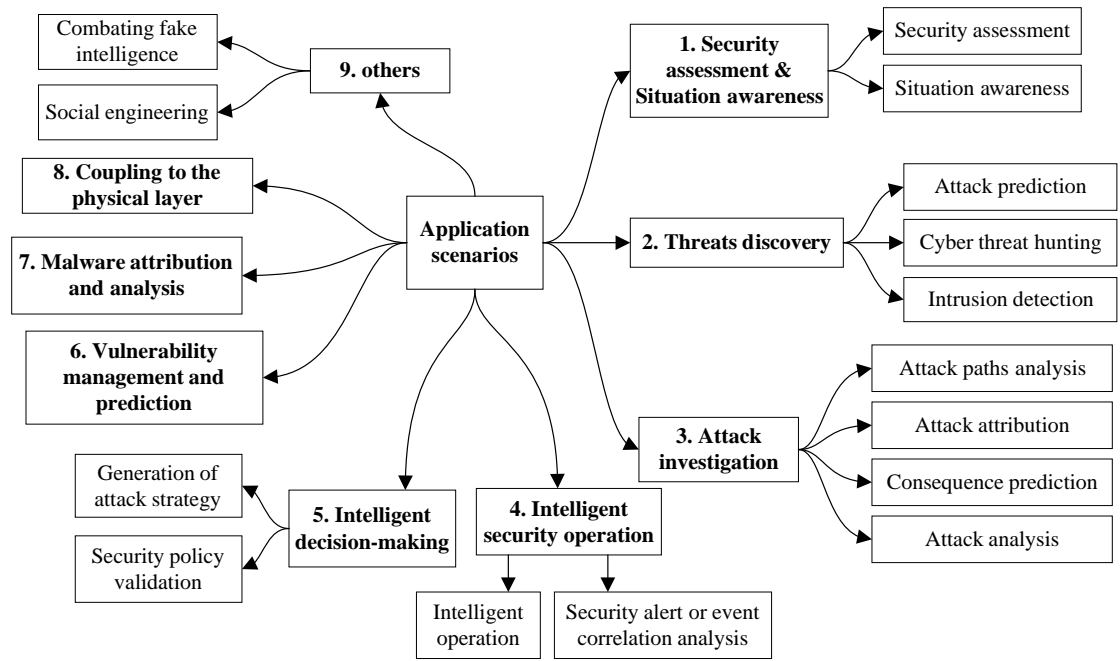

**Figure 4. Application scenarios of cyber security knowledge graph**

## 4.1. Situation awareness and security assessment

Assessing the overall security of an enterprise network as well as understanding its situation has become a demanding undertaking for human administrators as the combination of equipment and services deployed on enterprise networks has become increasingly complicated. A security administrator for an enterprise network has to combat these multi-stage and multi-host attack scenarios. CSKG can play a vital role in situation awareness and security assessment. MITRE proposed a situation awareness system CyGraph [86], which is mainly oriented toward network warfare task analysis, visual analysis, and knowledge management. As a KG with four layers: network infrastructure, security posture, cyber threats, mission readiness, CyGraph aims to analyze attack paths, predicts critical vulnerabilities, analyze intrusion alarm correlation, and interactive visual queries by bringing together isolated data and events into an ongoing overall picture. CyGraph provides some query-driven demonstration cases for stating its effectiveness but no corresponding datasets. Chen [87] combined an existing indicator system and situation detection model and proposes an attack situation detection scheme based on KG. It provides a novel feature for improving the accuracy of cyber security situation detection by abstracting the attack events (e.g., historical events, Internet news) as graph description. Wang [88] proposed a KG-based network security situation awareness model (KG-NSSA) to address the two classic problems: network attack scenario discovery and situation understanding. Different from the traditional alert-based attack scenario discovery approach, which is susceptible to a large number of redundancy and false positives, the scheme can effectively reflect the network attack scenario in the asset node situation with similarity estimating and attribute graph mining method. Yi et al. [89] utilized the domain knowledge-based reasoning method to realize automatic correlation analysis of multi-source intelligence and understand the satellite cyber situation.

### 4.1.1 security assessment

Wu et al. [90] proposed an innovative ontology and a graph-based approach for security assessment. The ontology, which may be instantiated for specific networks, is intended to represent security knowledge such as assets, vulnerabilities, and attacks in a standardized manner. Using the inference abilities of the ontological model, an efficient framework is proposed to generate attack graphs, identification of possible attacks caused by existing vulnerabilities, and assess network security. The final output of their method includes the attack graph in a visualized representation as well as descriptions of the properties set. To clearly show the progression of the attack, this paper provided a flow diagram of the attack process, which illustrates how an adversary can attack and compromise several objectives in the test network through multiple hosts and stages. This enables enterprise administrators to accomplish security risk assessment tasks and react to new threats. Inspired by the existing cybersecurity ontologies, such as STUCCO [30], UCO [31], Cyber Intelligence Ontology [91], Kiesling et al. [71] proposed a publicly available CSKG with concrete instance information and illustrated its applicability for security assessment by two SPARQL [92] query example scenarios. In order to assess the potential impact that a newly identified vulnerability may have, it will query which data assets might be exposed in the local system model based on linking organization-specific asset information to a continuously updated stream of known vulnerabilities. According to the application scenarios and security threat characteristics of power IOT terminals, Pang et al. [93] proposed a security assessment of power IOT terminals based on KG with three dimensions including terminal assets, vulnerability intelligence, and threat alarm. The method realizes the correlation analysis of the security monitoring data of the independent power

IoT terminal and generates the terminal threat index reflecting the terminal security status. This should have been a good attempt, unfortunately, the article did not give the details of the KG and method.

## 4.2. Threats Discovery

Even with advanced monitoring, sophisticated attackers can spend more than 100 days in a system before being detected [94]. Many factors ranging from alerts flooding to slow response-time, render existing techniques ineffective and unable to reduce the damage caused by these attacks. The CSKG could fill this gap with the power of knowledge representation and reasoning. This section presents the findings of the research about the application of CSKG, focusing on the three key themes including attack prediction, threat hunting, and intrusion detection. Before the cyberattack, security analyst hopes to spot signs of attacks ahead of time by attack prediction and threat hunting methods. During the attack, security administrators want to detect suspicious activities with intrusion detection technologies.

### 4.2.1 Attack prediction

Narayanan et al. [95] developed a cognitive system to detect cybersecurity events early by amalgamating information from traditional sensors, dynamic online textual sources, and KGs. The author extended UCO such that it can reason over the inputs from various network sensors like Snort, IDS, etc., and information from the cyber-kill chain. SWRL (Semantic Web Rule Language) was used to specify rules between entities. The aggregator module was designed to combine alerts into a reasoning model. They demonstrated its capabilities of detecting newer attacks by testing it against custom-built ransomware similar to WannaCry and showing the timeline of the attack performed and the actions from the system. Unfortunately, this study solely described the system's architecture and did not include any additional information or data. To solve the difficulty of attack prediction caused by the 0day vulnerability, a prediction method of 0day attack path based on cyber defense KG was proposed by Sun et al. [96]. The KG was generated from 3 aspects (i.e., threat, assets, and vulnerability), which supported transforming the attack prediction task to the link prediction problem of KG. According to this, a path ranking algorithm was applied to mine the potential 0day attack in the target system and construct the 0day attack graph. The experimental result showed that with the help of KG, the proposed method can improve the accuracy of 0day attack prediction. Moreover, utilizing the path ranking algorithm can also help to backtrack the reasons for predicting results so as to improve the explaining ability to predict.

### 4.2.2 Threat hunting

For the task of cyber threat hunting, a system was developed by Gao et al. [97] aiming at facilitating log-based cyber threat hunting by leveraging vast external threat knowledge provided by OSCTI [98]. The system is composed of two subsystems, i.e., a knowledge extraction pipeline for building threat behavior KG, and a query subsystem built upon system auditing, which could collect audit logging data across hosts. This research also provides a threat behavior query language (TBQL), and a query synthesis mechanism that automatically synthesizes a TBQL query, with event sequential information, from the threat behavior KG, which could be used for finding the matched system auditing records. Nevertheless, one of the limitations of this system is that it does not consider the attacks that are not captured by system auditing. Furthermore, existing methods often exhibit important limitations in terms of the interpretability, quantity, and relevance of the generated alerts.

### 4.2.3 Intrusion detection

Besides the intrusion detection methods mentioned by [99], the CSKG could also be constructive in detecting intrusion. Kiesling et al. [71] gave a query-based case to illustrate how alerts from the network intrusion detection system (NIDS) can be connected to the SEPSES cybersecurity KG in order to obtain a deeper understanding of potential threats and ongoing attacks. Chen et al. [100] proposed a DDoS attack detection method based on KG, which is mainly aimed at the DDoS attack on TCP traffic. The KG is used to express the communication process of TCP traffic between two hosts. Together with calculating the value for one-way transmission propensity, a threshold was set to determine whether the source host is the initiator of a DDoS attack. To comprehensively describe the DDoS attack, Liu et al. [101] constructed distributed DDoS attacks malicious behavior knowledge base, which contains two parts: a malicious traffic detection database and a network security knowledge base. The front one detects and classifies malicious traffic caused by DDoS attacks. The network security knowledge base is the core part of the malicious behavior knowledge base of DDoS attacks, including the network topology graph, malicious behavior traceability graph, malicious behavior feature graph, and traffic behavior KG. The network security knowledge base is responsible for data structure processing, malicious behavior KG construction, behavior reasoning, and feedback. In 2021, Garrido et al. [102] applied machine learning on KGs to detect unexpected activity in industrial automation systems integrating IT and OT elements. Using a readily available ontology [103], this study builds a KG via integrating three main sources of knowledge: information about the automation system, observations at the network level (e.g., connections between hosts), and observations at the application level (e.g., data access events). Inspired by KG completion methods, this study adopts a graph embedding algorithm to rank the likelihood of triple statements resulting from events observed. Experimentally, the suggested method produces intuitively well-calibrated and interpretable alarms in a variety of contexts, pointing to the potential benefits of relational machine learning on KG for intrusion detection. Although these results are produced on a reduced-scale prototype and without the help of CTI, the present research explores, for the first time, the synergistic combination of KG and industrial control systems.

## 4.3. Attack investigation

CSKG is fast becoming a key instrument in attack analysis. In the chapter that follows, we will discuss recent progress on how to apply KG into attack analysis from the following four aspects:

attack path analysis, attack attribution, consequence prediction, and attack analysis.

### 4.3.1 Attack path analysis

As mentioned above (in section 4.1), the CyGraph could query out potential attack paths based on the network environment. Similar to CyGraph, Neol et al. [104] illustrated a graph-based strategy with a novel attack graph model that merges a complex blend of network data, including topology, firewall policies, vulnerabilities, attack patterns, and intrusion alerts, through standardized languages for cyber security data. The author, furthermore, created a model that predicts possible attack paths based on network events (intrusion alerts, sensor logs, etc.). Correlating detected attack events with potential attack paths gives the best options for response, especially for protecting critical assets, and improves situational awareness, e.g., inferring missed attack steps and eliminating false positives. The resulting attack graph is stored in the Neo4j database [105] for query and visualization. Despite its efficiency of query and visualization for potential attack paths, the proposed KG still faces several disadvantages. Firstly, this research did not show us how to use KG to infer new knowledge. Secondly, the corresponding datasets, such as the input format of alerts data and firewall policies, were not demonstrated clearly. Finally, the OSCTI (the knowledge provided by METRE) is not reflected in the architecture. To extend the information on the attack path, Ye et al. [106] designed building a cyber attack KG with 4 types of entities including software, hardware, vulnerabilities, and attack entity. With the help of four kinds of attributions of attack entity (i.e., attack conditions, attack methods, success rate, and earnings), this research used KG to generate an attack path and improve the assessment of vulnerability rather than rely on CVSS score [107]. Thanks to the knowledge representation and information management ability of KG, the attack path could update local information based on multiple sources. To further increase efficiency, a graph-based strategy for determining the ideal penetration path is proposed, taking into account insider and unknown attacks. Wang et al. [37] defined a two-layer threat penetration graph (TLTPG), where the upper layer is a penetration graph of the network environment, and the lower layer is a penetration graph between any two hosts. The KG was used to describe the attack-related resources (e.g., software, vulnerability, ports in use, and privilege of a successful attack) of each host, which would be of great benefit not only to generate the penetration path between hosts but also to integrate collected information of 0-day vulnerability attack for unknown attack prediction. In the power networks, Chen et al. [108] generated the expansion attack graph for obtaining the maximum probability vulnerability path and providing the success rate of attacking the power grid and the loss. As was discussed in the "attack prediction" section, the KG could also be used to represent and generate 0day attack paths with the approach of link prediction and path ranking algorithm [96]. It comprehensively considered the existence, availability, and impact of vulnerability as well as the knowledge of attack intent, asset types.

### 4.3.2 Attack attribution

As a defender, we must be able to answer questions like who attacked me, where the attack point is, and what the attack path is in order to gain a competitive advantage in cyberwarfare. This step is known as attack attribution. The attack source, intermediate medium, and corresponding attack path can all be determined using attack attribution technologies, allowing for more tailored protection and countermeasure techniques to achieve active defense. As can be seen, attribution of attacks is a crucial step in the transition from passive to active defense. Based on an ontology with six dimensions, namely host asset, vulnerability, attack threat, evidence, location, strategy, and the relations between them, Zhu et al. [109] constructed a CSKG for space-ground integration information network. In addition, each dimension has several unique attributes and data sources. The research proposed an automated attributing framework for cyber-attack. Attack attribution could be performed from different perspectives by querying the established CSKG. As an example, given in the article, based on the host asset dimension, security personnel can query the KG to find out the vulnerable host asset that is suspected of being attacked, associated vulnerability, and attribution strategy in sequence. Then, the evidence and positions left by the attack could be found and the attacked host asset can be located by executing the corresponding attribution strategy. Xue et al. [110] analyzed the existing provenance graph construction technology based on causation in NSFOCUS's blog. The study introduced the provenance graph construction from three dimensions including terminal dimension, the perspective of Syslog and application log correlation, the association of network and terminal. The terminal perspective method focused on the relations of processes, files, and filenames in one isolated host and ignored the application log which was replenished by the second dimension. Moreover, the third level method extended the provenance graph from single host to multi-host network which could be improved by causal analysis for getting a complete attack process. However, this study did not consider the semantic context and OSCTI provided by CSKG.

### 4.3.3 Consequence prediction

Common software weaknesses, such as improper input validation, integer overflow, can harm system security directly or indirectly, causing adverse effects such as denial-of-service, execution of unauthorized code. Understanding the consequence of weakness becomes significant to assess the risk of a system and to take prompt response. In 2018, Han et al. [111] built a KG based on common weakness enumeration (CWE) [112], which contains rich information about weaknesses, including textual descriptions, common consequences, and relations between software weaknesses. The current CWE data does not support advanced reasoning tasks on software weaknesses, such as the prediction of missing relations and common consequences of CWEs. This research developed a translation-based, description-embodied knowledge representation learning method to embed both weaknesses and their relations in the KG into a semantic vector space. Following the vector embedding generation, extensive experiments were conducted to estimate the performance of KG in knowledge acquisition and inference tasks. In this study, CSKG could be exploited for three reasoning tasks, including CWE link prediction, CWE triple classification, and common consequence prediction. Datta et al. [113] transferred the consequence prediction problem to the classification task by introducing a dataset and building machine learning and natural language processing (NLP) models. The dataset of cyber-attacks consists of 93 diverse attacks and their descriptions, which are annotated with their technical and non-technical consequences. The idea is to enable security researchers to have tools at their disposal that makes it easier to communicate the attack consequences with various stakeholders who may have little to no cybersecurity expertise.

Additionally, with the proposed approach researchers' cognitive load can be reduced by automatically predicting the consequences of attacks in case new attacks are discovered.

### 4.3.4 Attack analysis.

Besides the above application, Qi et al. [114] built a CSKG, which includes two subgraphs: CSKG and scene KG, for attack analysis. The CSKG is the core graph representing the knowledge about vulnerabilities, attacks, assets and the relationships among them, which can be obtained from various vulnerability and attack analysis websites and can be updated gradually. Scene KG is an extended graph constructed based on node and connectivity information of the network involved in a specific attack. The input data of the whole analysis framework come from the data collection system and detection system. Using the CSKG, attack rule base, and spatiotemporal property constraints, composite attack chains are mined from multiple single attacks. With many alerts to investigate, cyber analysts often end up with alert fatigue, causing them to ignore a large number of alerts and miss true attack events. There is an observation that different attacks may share similar abstract strategies, regardless of the vulnerabilities exploited and payloads executed. Alsaheel et al. [115] presented ATLAS, a framework that constructs an end-to-end attack story from off-the-shelf audit logs based on a causal graph. ATLAS leverages a novel combination of causality analysis, NLP, and machine learning techniques to build a sequence-based model, which establishes key patterns of attack and non-attack behaviors from a causal graph. At inference time, given a threat alert event, an attack symptom node in a causal graph is identified. ATLAS then constructs a set of candidate sequences associated with the symptom node, uses the sequence-based model to identify nodes in a sequence that contribute to the attack, and unifies the identified attack nodes to construct an attack story.

6G-oriented network intelligence needs the support of knowledge from inside and outside the network. Therefore, Wang et al. [116] proposed a method to construct cyber-attack KG based on CAPEC [69] and CWE, which is implemented in the graph database Neo4j. This study only introduced two query-based application scenarios in detecting and responding to DDoS flood attacks and multi-stage attacks based on the query and display function provided by Neo4j, rather than based on the reasoning function of KG. This study just focused on the analysis and application of CAPEC and CVE, which did not cover the complete knowledge of cyber security.

## 4.4. Intelligent security operation

### 4.4.1 Intelligent operation

An AI-driven security operations framework was presented by Zhang et al. [84]. In this study, CSKG could support dynamic query and aggregation analysis of security data, improving the integrity of security data operation analysis. The KG is a unified data view, which can support the realization of multi-level technical capabilities such as subsequent risk perception, causal cognition, robust decision-making. And some challenges were discussed from the aspects of data, models, and semantic context. However, this research did not demonstrate the specific method.

The white paper [117] introduced the application scenarios of KG in the field of security operation from three aspects: attack profiling, attack path investigation and response mitigation strategy recommendation, as well as the challenges of intelligent operation. As a white paper, it aims to sort out directional content such as demand scenarios, application solutions, and technology prospects in this research field, but it will not involve technical details.

### 4.4.2 Security alert or event correlation analysis

Given the ever-evolving threat landscape, security researchers managing the security operation center (SOC) are often overloaded with numerous security incidents and, at the same time, trying to keep abreast with the latest threats in the wild. Effectively correlation analyzing large volumes of diverse alert or event data brings opportunities to identify issues before they become problems and to prevent future cyberattacks. Traditional methods usually store the different dimensions of security information in separate knowledge bases, which leads to the lack of synergies between the various dimensions. As illustrated by Xue [118], the main challenge faced by the application of cyber CSKG is that there is no direct connection between the KG based on abstract attack knowledge such as STIX 2.0 and the system and network logs that contain the behavior information. It is a semantic gap between them. For complex attacks, it is difficult to integrate all context information quickly to launch real-time and accurate analysis. The traditional rule-based association analysis needs to rely on expert knowledge to construct the attack scene which lacks the ability of reasoning automatically. Wang et al. [119] proposed an integrated security event correlation analysis system to solve the above problem. The system integrated the network infrastructure KG, vulnerability KG, cyber threat KG, and intrusion alert KG into the CSKG, and detailed the data source of each dimension. Following alert normalization and alert fusion, the alert verification was conducted by judging whether the vulnerabilities of one alert are in the host vulnerability set. Furthermore, the process of attack thread correlation analysis is based on the existing alerts to query the associated alerts, CVE items, and CAPEC items, which could be conducive to predicting the real purpose of attackers. In the author's thesis [120], rebuilding the scene of a series of alerts based on KG was introduced in detail. The author conducts an experiment on the DARPA 2000 dataset to evaluate the performance of the proposed framework by comparing the number of remaining alerts after correlation analysis. This research showed an example of the use of KG for correlation analysis. Qi et al. [121] considered that cyber-attacks have multiple attack steps, which are associated with alerts from intrusion detection systems (IDS). Based on this idea, an association analysis algorithm based on KG of cyber security attack events is proposed to display the attack scenario of an air-ground integrated network graphically. The CSKG contains 5 tuples: attack, event, alarm, relation, and rule. The association analysis was used by calculating the coincidence degree between the sequence of events collected and the sequence of events attacked in the KG. Nevertheless, due to a lack of comprehensive understanding of the space-ground integration network and the limitations of current experimental conditions, this paper merely used simulation experiments to verify the feasibility of the above algorithm. Logs manual investigation typically does not scale well and often leads to a lack of awareness and incomplete transparency about issues. To tackle this challenge, Ekelhart et al. [122] introduce a flexible framework for the automated construction of KGs from arbitrary raw log messages. By making log data amenable to semantic analysis, the workflow fills an important gap and opens up a wealth of data sources for KG building. As mentioned earlier in section 4.2, Garrido et al. [102] proposed the application of machine learning on KGs to improve the quality and relevance of IDS-generated alerts in modern industrial systems, increasing their usefulness for human operators.

## 4.5. Intelligent decision-making

The current cyber security assessment also relies on personal experience, and the level of intelligence is low. Improving the intelligence level of cyber security assessment is a problem that needs to be solved urgently. Based on the KG technology, it is worthwhile to study the decision model applicable to cyber security and improve the intelligence level of cyber security assessment. The aim of the section is to introduce several research cases of intelligent decision-making based on KG, such as the generation of attack strategy and security policy validation.

### 4.5.1. Generation of attack strategy

Analyzing the attack strategy from the perspective of an attacker can help determine existing security problems and provide targeted protection suggestions. Compared with the query-based method of CyGraph, a cyber-attack method recommendation algorithm based on KG was proposed by Ou et al. [123]. It contains a six-tuple KG construction schema based on 4 open databases (i.e., CVE, CWE, MSF, CAPEC), the collaborative filtering recommendation that describes difference relations between nodes by meta-path, a generator of recommendation list with calculating the correlation score of each path with node vector. In the second part, a recommendation algorithm for cyber attack entities is proposed by combining the method of machine learning feature extraction and the method of constructing heterogeneous information network meta-path. Based on this KG, intelligent search and recommendation of knowledge related to new threat intelligence can be achieved. Compared with the traditional content-based search recommendation method, this method is more accurate in predicting the weakness of vulnerabilities and can realize the prediction and recommendation of attack patterns based on the natural language description of vulnerabilities. Likewise, from the perspective of an attacker, Chen et al. [124] proposed a knowledge-driven attack strategy generation method to realize the combined exploitation of multiple vulnerabilities in the industrial control network. The method consists of a KG of vulnerability exploitation, a graphical representation of industrial control network, and knowledge reasoning rules. Searching for attack paths at the device level based on the attack process is a common idea among security experts. The process of formulating an attack strategy can be divided into two steps: The first step is analyzing multiple vulnerabilities on the current device-level nodes and correlating them based on the pre-conditions and consequences of exploitation. After formulating the exploit sequence of all the vulnerabilities in the device, connect the device-level nodes according to the access rules of the firewall and other protection devices to form a global attack strategy graphically. Currently, this proposed KG was applied to analyze multiple vulnerabilities on a small-scale industrial control network to generate attack paths. With the expansion of KG, such as supplementing with other threat intelligence, more and more attack strategies need to be generated, especially the most cost-effective attack strategy.

### 4.5.2. Security Policy Validation

Vassilev et al. [125] proposed a four-layer (i.e., ontological level, heuristic level, workflow level, and process-level) framework for logical analysis, threat intelligence analysis, and validation of security policies in cyber systems. The framework is validated using a set of scenarios describing the most common security threats in digital banking and a prototype of an event-driven engine for navigation through the intelligence graphs has been implemented. But this framework was developed specifically for application in digital banking and did not introduce used datasets.

## 4.6. Vulnerability management and prediction

This section presents several case studies that illustrate various CSKG analytic capabilities in vulnerability management and prediction. In security operation, managing, identifying, quantifying, and prioritizing vulnerabilities in one system is a key activity and a necessary precondition for threat mitigation and elimination and hence for the successful protection of valuable resources.

KG technologies provide an exciting opportunity to advance our knowledge of managing considerable vulnerability data by presenting them in a structured ontological format. Except for the using case in section 4.2.3, another using scenario of the SEPSES KG [71] is a query-based example to state how the KG can support security analysts by linking organization-specific asset information to a continuously updated stream of known vulnerabilities. Cybersecurity vulnerability ontology (CVO), a conceptual model for formal knowledge representation of the vulnerability management domain, was created by Syed et al. [126]. Additionally, in this research, they utilized the CVO to design a cyber intelligence alert (CIA) system that issues cyber alerts about impending vulnerabilities and countermeasures. At the practical level, its components include the vulnerability repository, social media intelligence extractor-tagger (SMIET), vulnerability mapper, RDF converter, CVO, the cyber intelligence ontology (CIO), and cyber alerts rules engine. Finally, this study gave the evaluation approaches, corresponding results, and examples in practice. Based on the industrial internet security vulnerabilities, an industrial CSKG was built and stored into Neo4j by Tao et al. [127], in order to analyze, query, and visualize from the dimension of temporal, spatial, and correlation.

In the face of actual attacks, CyGraph correlated intrusion alerts to known vulnerability paths and suggests the best courses of action for responding to attacks. CyGraph builds a query-based predictive model of possible attack paths and critical vulnerabilities. As previously stated, a CSKG based on CWE [111] could be used to predict missing relations and common consequences of CWEs by developing a translation-based, description-embodied knowledge representation learning method. To find hidden relationships among weaknesses, Qin et al.[40] proposed a query-based model for automatic analysis and reasoning. The reasoning flow of the sample CWE Chain was demonstrated based on vulnerability KG (VulKG) which contains the vulnerability data including NVD, CVE, CPE, and CWE. But the example could merely partially replace the analysis and labeling work of security experts under some specific scenarios, where the operator needs to know the query target previously. For effectively managing the sparse or inaccurate malware threat information, a malware KG called MalKG was established by Rastogi et al. [79], which is the first open-source automated KG for malware threat intelligence. Additionally, the provided MalKG dataset (i.e., MT40K) contains approximately 40,000 triples generated from 27,354 unique entities and 34 relations. The study also manually curated a benchmark KG dataset called MT3K, with 3,027 triples generated from 5,741 unique entities and 22 relations. It demonstrated the prediction capabilities of MalKG using two use cases in predicting new information. One of the application scenarios is predicting and sorting all the potential vulnerabilities or CVEs of the malware-impacted software system, by comprehensive utilization of information from the network environment, malware, and KG. As has already been noted in [124], a KG about vulnerability exploitation was established by integrating and extracting multi-dimensional domain knowledge. By occupying each device-level node, attack strategies based on KG improve the performance in comprehensive vulnerability exploitation and flexible response. The feasibility of the method was demonstrated through an industrial network example. Similarly, in Wang's study [128] chain reasoning and confidence calculation were also used for supporting vulnerability detection and finding latent relations between CWEs. At the end of this research, similarity matching based on a source code level graph is used for judging the similarity between target node and node in the vulnerability database, which provides new insights into vulnerability mining. Wang et al.[129] extended the relations in vulnerability KG by identifying the alternative vulnerability with similar consequences.

## 4.7. Malware attribution and analysis

This section aims to discuss how to utilize the KG to analyze and attribute the malware. Najafi et al. [130] proposed a graph-based malware rank inference algorithm, named MalRank, which was designed to infer a node maliciousness score based on its associations to other entities presented in the KG, e.g., shared IP ranges or name servers. This essay presented a KG that models global associations among entities observed in proxy and IDS logs, enriched with related open-source intelligence (OSINT) and CTI. The authors formulate threat detection in the security information and event management (SIEM) environment as a large-scale graph inference problem. After a series of experiments on real-world data captured from a global enterprise's SIEM, it showed that MalRank maintains a high detection rate outperforming its predecessor, belief propagation, both in terms of accuracy and efficiency. Furthermore, it showed that this approach is effective in identifying previously unknown malicious entities such as malicious domain names and IP addresses. Besides the application scenarios reported earlier, MalKG [131] could also be implemented in the malware attribution scenario [79]. For example, given a newly discovered malware attack on one system, the analyst needs to build a fingerprint of the malware's origination by assembling sufficient features, such as author, campaign, and others. The goal of MalKG is to automate the prediction of these features associated with a given malware, for instance, the newly discovered malware may share similarities with a disclosed malware linked to a certain APT group. As reported in the white paper [117], the profiling and automatic attribution of APT attacker gangs can be realized through the extraction of key elements of threat intelligence and dynamic behavioral reasoning. The key solution lies in how establishing a unified language to describe the behavior and characteristics of different APT organizations, as well as build a knowledge base about APT organizations. But the white paper did not disclose the details of related research.

## 4.8. Connection to the physical system

The CSKG uses big data analysis and graph mining technology to deeply analyze the coupling relationship between the information layer and the physical layer in the modern industrial control system, and realize the intelligence of "decision making, risk prediction, accident analysis, attack identification" and other capabilities assisted and automated processing. This section attempts to provide a summary of the literature relating to how to combine the CSKG with a physical network environment.

To illustrate various cybersecurity analytic capabilities in MITRE's situation awareness system CyGraph [86] and [104], they presented a simple network architecture for the case study. The architecture shows the underlying connectivity among hosts, switches, routers, and firewalls. The internal network is segmented into three protection domains (DMZ, mission client workstations, and data center). The external firewall protects the internal network from the outside, and the internal firewall protects the critical data-center servers. Based on the information on the network topology, firewall rules, and vulnerability scan results, the KG was built. To verify the effectiveness of the proposed method, a typical internal network architecture model with six types of elements is introduced in this paper[106]. In this architecture, the firewall isolates the Internet from the intranet router. FTP server, host1, and host2 are directly connected to the router. Host1 and host2 can access the FTP server. The database server is connected to the FTP server to receive and respond to requests from the FTP server. For the purpose of generating penetration paths, the essay [37] designed an illustrative network example. The network contains a host on Internet, a DMZ area, and three subnets. There is a web server in the DMZ area. Subnet1 has two devices (i.e., one Pad and a host), which can be connected to the Internet. Subnet2 has two hosts and cannot connect to the Internet. Subnet3 includes three servers including a print server, file server, and data server. The attacker is a host on the Internet. It also considered the potential connection between subnet1 and subnet2 with USB. Similar to the architecture above, an experiment network environment was designed in the paper [96]. In this study, the experiment network contains two subnets protected by two firewalls separately, and one DMZ also with a firewall. The two subnets connect the Internet through the DMZ. Each part of the network has different assets, such as an email server and a web server are in the DMZ, two hosts and a file server are part of subnet1, and an application server is connected with subnet2. Building a suitable experiment network environment could be beneficial to demonstrate the effect of approaches and reproduce the attack and defensive process.

Despite the various traditional network architectures, it is also important to research the security of industrial control systems with a suitable experiment network. In the research of [102], a hardware prototype was described for evaluation, following the design of modern industrial systems integrating IT and OT elements. The automation side is equipped with a Siemens S7-1500 PLC connected to peripherals via an industrial network. These peripherals include a drive subsystem controlling the motion of a conveyor belt, an industrial camera, a human-machine interface (HMI), and a distributed I/O subsystem with modules interfacing with various sensors for object positioning and other measurements. The PLC exposes values reported by these sensors as well as information about the state of the system through an OPC-UA server. So, the PLC connects to 2 edge computing servers. Thereafter, the network with main traffic flows was also displayed in this paper. In order to sense and measure the security risks and threats of massive power IoT terminals in real-time, a security threat assessment for power IOT terminals based on KG was proposed [93]. But this article did not describe a suitable network for evaluation. As analyzed previously, Chen et al. [124] applied the domain KG to analyze multiple vulnerabilities in the industrial control network to generate attack strategies. The topology of the target network is composed of the Internet, two firewalls, one router, an enterprise network, and an industrial ethernet. One firewall is used to protect all assets of the local network. The other one is situated between the enterprise network and industrial ethernet. The route is between the first firewall and the enterprise network, followed by the second firewall and industrial ethernet. The assets of the enterprise network include a web server, admin host, and printer. Some peripherals, such as an HMI, a data server, a workstation, and three PLCs with different end-effector devices (e.g., valve, flowmeter), are connected to the industrial ethernet. The attacker is a certain host from the Internet and the target of the attack is the PLCs.

Based on the above analysis and related research, this paper sorts out a general experiment network architecture (as shown in Figure 5) to demonstrate the effect of potential security investigation approaches. This general network mainly contains four parts including DMZ, a subnet connected to the Internet via a router, a subnet connected to DMZ, and an industrial control network connected to DMZ. Each subnet is isolated by a firewall. And the attackers usually start their offensive action from the Internet. A researcher could utilize it to adapt to a complex network by modifying or adding some devices, extending the subnets, or changing the connection mechanism. The network topology expresses the network environment. In addition, it should also include the software and hardware installed on each node, security protection measures, and existing vulnerabilities.

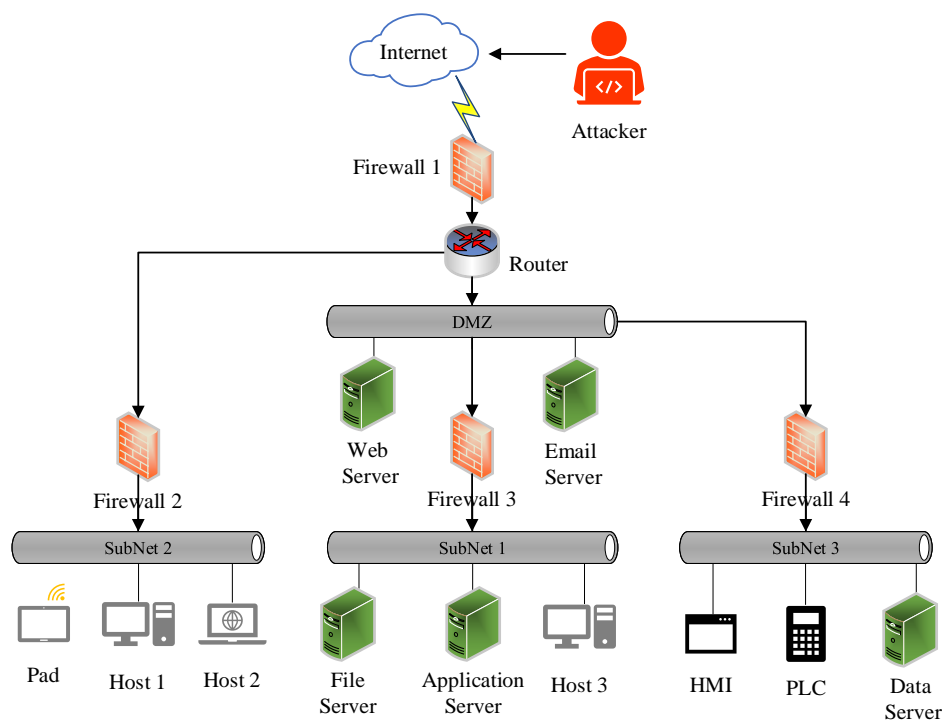

Figure 5 A general experiment network architecture

## 4.9. Other reasoning tasks

So far, this paper has focused on several dominating application scenarios. The following section will discuss some fresh applications, such as social engineering attack analysis, fake cyber threat intelligence identification, and so on.

Succinctly, social engineering is a type of attack wherein the attacker exploits human vulnerability through social interaction to breach cyberspace security [132]. Social engineering has posed a serious threat to cyberspace security. To protect against social engineering attacks, Wang et al. [133] develops an ontology of social engineering in cybersecurity and conducts an ontology evaluation method by its applications. The ontology defines 11 concepts of core entities that significantly constitute or affect the social engineering domain, together with 22 kinds of relations. It provides a formal and explicit knowledge schema to understand, analyze, reuse and share domain knowledge of social engineering. Furthermore, the KG was built based on 15 social engineering attack incidents and scenarios, and it was thoroughly evaluated using 7 query-based application examples (in 6 analysis patterns) comprehensively.

Today there is a significant amount of fake cybersecurity intelligence on the internet. To filter out such information, Mitra et al. [134] build a system to capture the provenance information and represent it along with the captured CTI. Together with enhancing the exiting CSKG model to incorporate intelligence provenance, this study fused provenance graphs with CSKG. The reasoning capabilities of CSKG enforce rules that help in preserving credible information and discarding the rest. Additionally, classes capturing the provenance can be added to the schema of the CSKG which can give us more information about the source of the data. However, the details and datasets of this novel KG were not given.

Apart from that, Xiao et al. [135] developed a KG embedding method to predict within-type and across-type relations of software security entities. Finding such missing relationships among existing entities helps analysts enrich software security knowledge. But this CSKG is not open-source so we could not read the details of it. In addition, the mentioned white paper [117] reported several other application scenarios of CSKG technology as well as its two classical reasoning methods. Despite the fact that there was some limitation in stating the adequate details, the application scenarios, such as ATT&CK threat modeling, APT threat hunting, intelligent security operation, cyberspace surveying and mapping, supply chain security, cyber-physical system protection, were outlined and forecasted by the white paper. There are two broad categories of reasoning technologies based on CSKG: relational reasoning based on graph representation learning and multi-relational reasoning methods based on neural networks.

# 5 Discussion and research opportunities

At present, some achievements have been made in the application of KGs in multiple knowledge-driven cyber security tasks, but in general, it is still in its infancy. In this section, we first give a brief summary of these methods to identify the gap and then propose research opportunities for relevant aspects of CSKG.

**(1) Open-source dataset construction**

In reviewing the literature, no open-source data was found on the perfect solution to all the problems. For the task of CSKG construction, further research should be carried out on building a dataset that could cover all the dimensions such as the knowledge data from MITRE, the CTI obtained from open sources, the environment data (e.g., assets, attributions, and their topology), and behavior data (such as network alerts, terminal alerts, and logs), as shown in Figure 6.

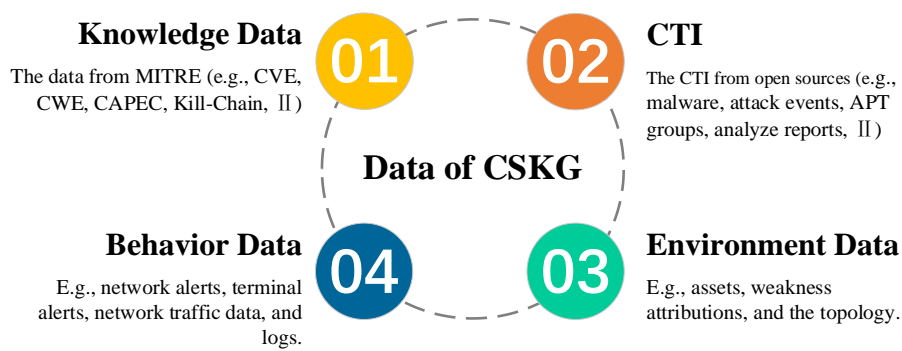

**Figure 6 The data for the cyber security knowledge graph construction**

The adequate annotated cyber security datasets are indispensable for training or validating the IE models, even for the pre-training language model or the prompt-based language models. However, existing datasets could not support this task well because of several drawbacks: firstly, most of them are designed for only one information extraction task (i.e., entity extraction) rarely for two IE tasks; secondly, because of different self-designed ontologies and different research targets, the entity and relation types are various; thirdly, the existing datasets are in a single language (i.e., English), which could not satisfy the requirement of multi-language; finally, annotating the corpus manually remains the primary way to offer initial data for the model in the vertical domain.

Further research should be undertaken to investigate the new multi-language cyber security IE dataset building based on comprehension and reliable data sources. This potential dataset should be annotated in a standard format and with a statement document. In the aspect of the annotated method, to lessen the reliance on annotated vertical corpus, semi-supervised or unsupervised extraction approaches, as well as prompt-based generating methods, can be investigated.

**(2) The construction of a dynamic cyber security knowledge graph**

There are mature frameworks for reference to construct the knowledge graph. Both top-down construction methods [136] and bottom-up construction methods [137] can be used to build large-scale knowledge bases. In the area of cyber security, the former one is more popular (i.e., design a cyber security ontology schema firstly, then extract the knowledge required by the schema from the corpus), which relies heavily on expert knowledge. The automatic ontology construction technology (also known as ontology learning) should still be considered necessary for the timely collection of emerging knowledge during the process of ontology update.

Conventional knowledge graph mainly focuses on the entities, their relations, attributions, etc., which are relatively deterministic and static knowledge. With the development of KG research and the demand for field applications, event knowledge and dynamic knowledge, such as temporal information, conditional relationships, causal information, and event subordination relationships, will inevitably be included. Considerably more work will need to be done to represent the cyber security events knowledge and support relevant logical reasoning by building a cyber security event temporal knowledge graph.

**(3) The application scenarios of the cyber security knowledge graph**

Although the construction technologies of CSKG are stable, there is still no unified open-source KG that is accepted by everyone. Despite their value and practicality, KGs usually suffer from incompleteness, redundancy, and ambiguity that might translate to uninformative query results. As the result of different application demands of various scenarios, researchers have to rebuild a new knowledge graph every time. This survey has made a comprehensive review of the application scenarios of CSKG, but at present, the CSKG function proposed above mainly remains on the query and display functions provided by Neo4j. This does not fully exploit the KG's potential to automate reasoning. Therefore, it is still not clear how to use it to solve some practical problems in the cyber security domain. KG completion is just one of the many applications of knowledge reasoning technology. To get a new understanding, additional exploration and study based on reasoning technology should be conducted.

The semantic gap between CSKG and logs is the key to restricting the application of CSKG to attack path investigation. By supplementing relevant knowledge, this semantic gap can be filled, and the semantic association between CSKG and log can be realized. The most important work in the future is to improve the interaction between the CSKG and the network internal knowledge, particularly the cyber-physical system, as well as to apply the KG's automated reasoning ability and association analysis ability to uncover risks and network situational awareness.

**(4) The evaluation criterion of the cyber security knowledge graph**

Although it is still in its infancy, potential applications of CSKG are found both in defensive and offensive scenarios across most cybersecurity functions. Currently, there are no established evaluation standards for the KG. The accuracy, precision, and F1 value are frequently used by researchers to assess the information extraction model. Hits@n, Mean Rank (MR), and Mean Reciprocal Rank (MRR) is used for assessing the triple prediction model's reasoning ability and use the query-based cases to demonstrate the KG's query and visualization capabilities. These aren't ideal for a comprehensive analysis of a knowledge graph. For example, we cannot claim that the KG may be used in certain situations to demonstrate that it is superior to other KG. Accordingly, future studies on proposing the evaluation standards for CSKG are therefore recommended.

# 6 Conclusion

In this review, we have provided a critical overview of the various works on the application scenarios of the cyber security knowledge graph. To begin, this paper introduces a brief overview of the background, concepts, and construction technologies of the cyber security knowledge graph. Then, several open-source datasets that are available for building cyber security knowledge graph and the information extraction task, and their drawbacks are illustrated. In the fourth part of this paper, we carried out a comparative study of the different works that elaborate on the recent progress in the application scenarios of CSKG. A novel comprehensive classification framework was developed for describing the related works from nine main aspects and eighteen subclasses. Finally, based on the discussion of shortcomings of existing research, future research directions have been prospected.

Security managers can use KG to intuitively understand security intelligence, network situation, entity relationships, and then discover the attributes of security entities, which could serve as a foundation for understanding cyber security knowledge, analyzing cyber security data, and discovering attack patterns and abnormal characteristics related to cyber-attacks. It is hoped that this research will contribute to a deeper understanding of how to apply cyber security knowledge graphs in industrial practice.